\documentclass[12pt]{article}
\usepackage{axodraw}

\catcode`@=12
\begin{document}
\thispagestyle{empty}
\begin{flushright} 
UCRHEP-T369\\ 
January 2004\
\end{flushright}
\vspace{0.5in}
\begin{center}
{\LARGE	\bf Quarks and Leptons in a Hexagonal Chain\\}
\vspace{1.5in}
{\bf Ernest Ma\\}
\vspace{0.2in}
{\sl Physics Department, University of California, Riverside, 
California 92521\\}
\vspace{1.5in}
\end{center}
\begin{abstract}\
The seemingly disparate notions of chiral color and quark-lepton 
nonuniversality are combined, and shown to be essential to each other as 
part of an underlying (and unifying) larger symmetry, i.e. supersymmetric 
$SU(3)^6$. Both phenomena are accessible experimentally at the TeV energy 
scale.
\end{abstract}
\newpage
\baselineskip 24pt

In the Standard Model of quarks and leptons, the electric charge has two 
components, i.e.
\begin{equation}
Q = I_{3L} + {1 \over 2} Y,
\end{equation}
where $I_{3L}$ comes from $SU(2)_L$ and $Y$ from $U(1)_Y$.  If the gauge 
group is extended to include $SU(2)_R$, then there are two possible 
decompositions of the electric charge.  One is based on \cite{ps74} 
$SU(4)_C \to SU(3)_C 
\times U(1)_{B-L}$, i.e.
\begin{equation}
Q = I_{3L} + I_{3R} + {1 \over 2} (B-L).
\end{equation}
The other is based on $SU(3)_L \to SU(2)_L \times U(1)_{Y_L}$ and $SU(3)_R 
\to SU(2)_R \times U(1)_{Y_R}$, i.e.
\begin{equation}
Q = I_{3L} + I_{3R} - {1 \over 2} Y_L - {1 \over 2} Y_R.
\end{equation}
[The minus signs in the above expression are due to a convention which will 
become clear later.]  Whereas Eq.~(2) is indicative of $SO(10)$ as the 
unification group, Eq.~(3) is indicative of \cite{dgg84} $SU(3)_C \times 
SU(3)_L \times SU(3)_R$.  However, the two are in fact equivalent if 
considered as subgroups of $E_6$.

Using Eq.~(2), under $SU(2)_L \times SU(2)_R \times U(1)_{B-L}$, the 
quarks and leptons transform as
\begin{equation}
q = (u,d)_L \sim (2,1,{1\over 6}), ~~~ q^c = (d^c,u^c)_L \sim (1,2,
-{1 \over 6}),
\end{equation}
\begin{equation}
l = (\nu,e)_L \sim (2,1,-{1 \over 2}), ~~~ l^c = (e^c,\nu^c)_L \sim (1,2,
{1 \over 2}).
\end{equation}
They are different in their electric charges because they have different $B-L$ 
values.  Using Eq.~(3), under $SU(2)_L \times SU(2)_R \times U(1)_{Y_L} 
\times U(1)_{Y_R}$,
\begin{equation}
q \sim (2,1,{1 \over 6},0), ~~~ q^c \sim (1,2,0,-{1 \over 6}),
\end{equation}
\begin{equation}
l \sim (2,1,-{1 \over 6},-{1 \over 3}), ~~~ l^c \sim (1,2,{1 \over 3},
{1 \over 6}).
\end{equation}
Here the differences come from the fact that the leptons belong to the 
$(1,3,3^*)$ representation of $SU(3)_C \times SU(3)_L \times SU(3)_R$, 
whereas $q$ and $q^c$ belong to the $(3,3^*,1)$ and $3^*,1,3)$ 
representations respectively.  Note that $(Y_L,Y_R) = (-1/3,0), (0,-1/3), 
(1/3,2/3), (-1/3,-2/3)$ respectively for $q,q^c,l,l^c$ in these 
representations.  This explains the minus signs in Eq.~(3).

At this point, a curious fact must have already been noticed, i.e. 
the electric charge has two components in Eq.~(1), three in Eq.~(2), and 
four in Eq.~(3).  How about five or more?  Using the idea of a separate 
$SU(3)_l$ for leptons \cite{fl90}, which results in a successful 
nonsupersymmetric $SU(3)^4$ model \cite{bmw03}, the electric charge may 
indeed have five components, i.e.
\begin{equation}
Q = I_{3L} + I_{3R} - {1 \over 2} Y_L - {1 \over 2} Y_R - {1 \over 2} Y_l,
\end{equation}
where $Y_l$ comes from $SU(3)_l \to SU(2)_l \times U(1)_{Y_l}$.  In this case,
\begin{equation}
q \sim (2,1,{1 \over 6},0, 0), ~~~ q^c \sim (1,2,0,-{1 \over 6},0),
\end{equation}
\begin{equation}
l \sim (2,1,-{1 \over 6},0,-{1 \over 3}), ~~~ l^c \sim (1,2,0,{1 \over 6},
{1 \over 3}).
\end{equation}
Going back to Eq.~(3), it is also clear that quarks and leptons may belong 
to \underline {different} $SU(3)_L$'s and $SU(3)_R$'s, so that the electric 
charge has eight components, i.e.
\begin{equation}
Q = (I_3)_{qL} + (I_3)_{qR} - {1 \over 2} Y_{qL} - {1 \over 2} Y_{qR} + 
(I_3)_{lL} + (I_3)_{lR} - {1 \over 2} Y_{l_L} - {1 \over 2} Y_{lR}.
\end{equation}

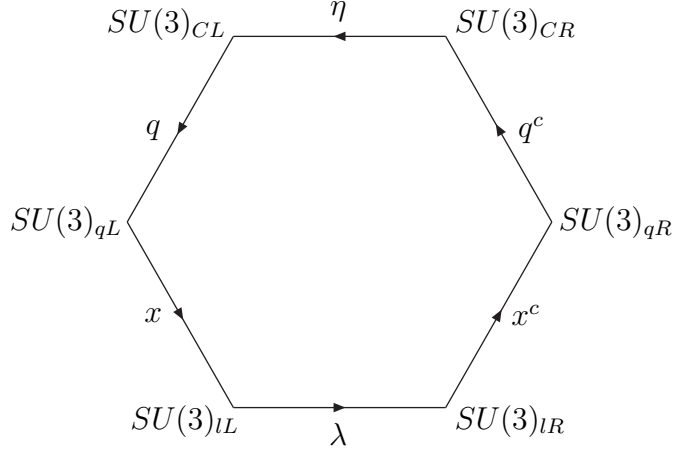
\begin{figure}[htb]
\begin{center}
\begin{picture}(270,150)(0,0)

\ArrowLine(100,0)(180,0)
\ArrowLine(60,70)(100,0)
\ArrowLine(100,140)(60,70)
\ArrowLine(180,140)(100,140)
\ArrowLine(220,70)(180,140)
\ArrowLine(180,0)(220,70)

\Text(70,105)[]{$q$}            \Text(75,145)[]{$SU(3)_{CL}$}
\Text(213,105)[]{$q^c$}         \Text(207,145)[]{$SU(3)_{CR}$}
\Text(70,35)[]{$x$}             \Text(37,70)[]{$SU(3)_{qL}$}
\Text(211,35)[]{$x^c$}          \Text(245,70)[]{$SU(3)_{qR}$}
\Text(140,-10)[]{$\lambda$}     \Text(82,-5)[]{$SU(3)_{lL}$}
\Text(140,150)[]{$\eta$}        \Text(205,-5)[]{$SU(3)_{lR}$}


\end{picture}
\end{center}
\caption{Moose diagram of quarks and leptons in $[SU(3)]^6$.}
\end{figure}
Combining this notion of quark-lepton nonuniversality \cite{gjs89,mr90,lm03} 
with that of chiral color \cite{fg87}, the group $SU(3)^6$ is then obtained.  
Note that this is very different from the previously proposed \cite{su4^6} 
nonsupersymmetric $SU(4)^6$ model which predicted a value of $\sin^2 \theta_W$ 
very far away from the present data.  A good way of displaying the structure 
of this symmetry is again a hexagonal ``moose'' diagram \cite{moose} 
(Fig.~1) with the assignments
\begin{eqnarray}
q &\sim& (3,3^*,1,1,1,1), \\ 
x &\sim& (1,3,3^*,1,1,1), \\ 
\lambda &\sim& (1,1,3,3^*,1,1), \\ 
x^c &\sim& (1,1,1,3,3^*,1), \\ 
q^c &\sim& (1,1,1,1,3,3^*), \\ 
\eta &\sim& (3^*,1,1,1,1,3),
\end{eqnarray}
under $SU(3)_{CL} \times SU(3)_{qL} \times SU(3)_{lL} \times SU(3)_{lR} \times 
SU(3)_{qR} \times SU(3)_{CR}$.  It reduces to the well-known $SU(3)_C \times 
SU(3)_L \times SU(3)_R$ model if the $x,x^c,\eta$ links are all contracted. 
It is also the natural \underline {anomaly-free} extension of chiral color 
(the $\eta$ link) and quark-lepton nonuniversality (the $x,x^c$ links).

The particle content of this model is given by
\begin{equation}
q = \pmatrix {d & u & h \cr d & u & h \cr d & u & h}, ~~~ 
q^c = \pmatrix {d^c & d^c & d^c \cr u^c & u^c & u^c \cr h^c & h^c & h^c}, ~~~ 
\lambda = \pmatrix {N & E^c & \nu \cr E & N^c & e \cr \nu^c & e^c & S},
\end{equation}
where the rows denote $(I_3,Y) = (1/2,1/3), (-1/2,1/3), (0,-2/3)$, and the 
columns denote $(I_3,Y) = (-1/2,-1/3), (1/2,-1/3), (0,2/3)$, with $x$ and 
$x^c$ having the same electric charge assignments as $\lambda$, and all 
the components of $\eta$ are neutral.  The doubling of $SU(3)^3$ to 
$SU(3)^6$ also allows the six gauge couplings to unify with $\sin^2 \theta_W$ 
equal to the canonical 3/8 at the unification scale.  To check this, 
consider the contributions of $q,x,\lambda,x^c,q^c,\eta$ to $\sum I_{3L}^2$ 
and $\sum Q^2$:
\begin{equation}
\sin^2 \theta_W = {\sum I_{3L}^2 \over \sum Q^2} = {{3 \over 2} + 3 + 
{3 \over 2} + 0 + 0 + 0 \over 2 + 4 + 4 + 4 + 2 + 0} = {6 \over 16} = {3 \over 
8},
\end{equation}
as expected.  Note that in the $SU(3)^4$ model \cite{bmw03} of leptonic color, 
$\sin^2 \theta_W = 1/3$ at the unification scale.

There is another reason for choosing $SU(3)^6$.  If chiral color is invoked 
without separate $SU(2)_L$ gauge groups for quarks and leptons, then an 
$SU(3)^4$ model with $\sin^2 \theta_W = 3/8$ is possible.  However, because
\begin{equation}
{1 \over \alpha_s} = {1 \over \alpha_{CL}} + {1 \over \alpha_{CR}},
\end{equation}
$\alpha_s$ would be wrong by at least a factor of two.  As it is, $SU(3)^6$ 
allows the intriguing possibility that \underline {both} chiral color and 
quark-lepton nonuniversality may exist at experimentally accessible energies, 
as shown below.

Let the neutral scalar components of the supermultiplets corresponding to 
$x^c_{11}$, $x^c_{22}$, $x^c_{13}$, $x^c_{31}$, $x^c_{33}$, $x_{33}$, and 
$\lambda_{33}$ acquire large vacuum expectation values, then $SU(3)^6$ is 
broken down to
\begin{equation}
SU(3)_{CL} \times SU(3)_{CR} \times SU(2)_{qL} \times SU(2)_{lL} \times 
U(1)_Y.
\end{equation}
This symmetry embodies both the notions of chiral color and quark-lepton 
nonuniversality and is assumed to be valid down to $M_S$, the supersymmetry 
breaking scale.  The particles which remain massless between $M_S$ and $M_U$, 
the unification scale, are assumed to be three copies of $q$, $q^c$, and 
$\eta$, three copies of all the components of $x$ except $x_{33}$, three 
copies of $(\nu,e)$ and $(e^c,\nu^c)$ in $\lambda$, but only one copy of 
the $(N,E;E^c,N^c)$ bidoublet in $\lambda$.  This particle content 
generalizes that of the MSSM (Minimal Supersymmetric Standard Model), where 
there are three families of quarks and leptons, but only one pair of Higgs 
doublets.  The transformation properties of these particles with respect to 
this symmetry are then given by
\begin{eqnarray}
&& (u,d) \sim (3,1,2,1,1/6), ~~~ h \sim (3,1,1,1,-1/3), ~~~ \eta \sim 
(3^*,3,1,1,0), \\ 
&& d^c \sim (1,3^*,1,1,1/3), ~~~ u^c \sim (1,3^*,1,1,-2/3), ~~~ h^c \sim 
(1,3^*,1,1,1/3), \\ 
&& (\nu,e) \sim (1,1,1,2,-1/2), ~~~ e^c \sim (1,1,1,1,1), ~~~ \nu^c \sim 
(1,1,1,1,0), \\ 
&& (\nu_x,e_x) \sim (1,1,2,1,-1/2), ~~~ (e^c_x,\nu^c_x) \sim (1,1,1,2,1/2), \\ 
&& (N_x, E_x; E^c_x, N^c_x) \sim (1,1,2,2,0), \\ 
&& (N,E) \sim (1,1,1,2,-1/2), ~~~ (E^c,N^c) \sim (1,1,1,2,1/2).
\end{eqnarray}
Above $M_S$, the theory is supersymmetric and all these supermultiplets  
contribute to the running of the five gauge coupings of Eq.~(21).  The 
one-loop renormalization-group equations are given by 
\begin{equation}
{1 \over \alpha_i (M_S)} - {1 \over \alpha_i (M_U)} = {b_i \over 2 \pi} 
\ln {M_U \over M_S},
\end{equation}
where
\begin{eqnarray}
SU(3)_{CL} &:& b_{CL} = -9 + 3N_f = 0, \\ 
SU(3)_{CR} &:& b_{CR} = -9 + 3N_f = 0, \\ 
SU(2)_{qL} &:& b_{qL} = -6 + 3N_f = 3, \\ 
SU(2)_{lL} &:& b_{lL} = -6 + 2N_f + 1 = 1, \\ 
U(1)_Y &:& b_Y = 5N_f + 1 = 16,
\end{eqnarray}
where $N_f =3$ is the number of families.

At $M_S$, in addition to the breaking of supersymmetry, assume as well 
that $SU(3)_{CL} \times SU(3)_{CR}$ is broken to $SU(3)_C$ through the 
vacuum expectation values of the diagonal elements of $\eta$, and $SU(2)_{qL} 
\times SU(2)_{lL}$ is broken to $SU(2)_L$ through the vacuum expectation 
values of $N_x$ and $N^c_x$.  The boundary conditions at $M_S$ are
\begin{eqnarray}
{1 \over \alpha_s (M_S)} &=& {1 \over \alpha_{CL} (M_S)} + {1 \over 
\alpha_{CR} (M_S)}, \\ 
{1 \over \alpha_2 (M_S)} &=& {1 \over \alpha_{qL} (M_S)} + 
{1 \over \alpha_{lL} (M_S)}.
\end{eqnarray}
Below $M_S$, the particle content becomes 
that of the Standard Model, but with two Higgs doublets, i.e.
\begin{eqnarray}
SU(3)_C &:& b_s = -11 + (4/3) N_f = -7, \\ 
SU(2)_L &:& b_2 = -22/3 + (4/3) N_f + 1/3 = -3, \\ 
U(1)_Y &:& b_Y = (20/9) N_f + 1/3 = 7.
\end{eqnarray}
At $M_U$, all six gauge couplings are assumed equal.  Using $\sin^2 \theta_W 
(M_U) = 3/8$, this means that
\begin{equation}
{3 \over 5 \alpha_Y (M_U)} = {1 \over \alpha_2 (M_U)} = {2 \over \alpha_U}.
\end{equation}
Putting all these together, the constraints on the gauge couplings of this 
model are then given by
\begin{eqnarray}
{1 \over \alpha_s (M_Z)} &=& - {7 \over 2 \pi} \ln {M_S \over M_Z} + {2 \over 
\alpha_U}, \\ 
{1 \over \alpha_2 (M_Z)} &=& - {3 \over 2 \pi} \ln {M_S \over M_Z} + {2 \over 
\pi} \ln {M_U \over M_S} + {2 \over \alpha_U}, \\ 
{3 \over 5 \alpha_Y (M_Z)} &=& {21 \over 10 \pi} \ln {M_S \over M_Z}  + {24 
\over 5 \pi} \ln {M_U \over M_S} + {2 \over \alpha_U}.
\end{eqnarray}
These equations are easily solved for $\alpha_s (M_Z)$ and $M_U/M_Z$ in terms 
of $\alpha_2 (M_Z)$, $\alpha_Y (M_Z)$, and $M_S/M_Z$, i.e.
\begin{eqnarray}
{1 \over \alpha_s (M_Z)} &=& {3 \over 7} \left( {4 \over \alpha_2 (M_Z)} - 
{1 \over \alpha_Y (M_Z)} \right) + {4 \over 7 \pi} \ln {M_S \over M_Z}, \\ 
\ln {M_U \over M_Z} &=& {\pi \over 14} \left( {3 \over \alpha_Y (M_Z)} - 
{5 \over \alpha_2 (M_Z)} \right) - {2 \over 7} \ln {M_S \over M_Z}.
\end{eqnarray}
Using the input \cite{pdg}
\begin{eqnarray}
\alpha_2 (M_Z) &=&  (\sqrt 2/\pi) G_F M_W^2 = 0.0340, \\ 
\alpha_Y (M_Z) &=& \alpha_2 (M_Z) \tan^2 \theta_W = 0.0102,
\end{eqnarray}
the value of $\alpha_s (M_Z)$ is predicted to be in the range 0.119 to 0.115 
for $M_S/M_Z$ in the range 1 to 5, as shown in Table I, compared to the 
experimental world average \cite{pdg} of $0.117 \pm 0.002$.  The value of 
$M_U$ is of order $10^{16}$ GeV, in good agreement with the usual theoretical 
expectation.  Thus a \underline {new} and remarkably 
\underline {successful} model of grand unification is obtained.  It is also 
experimentally \underline {verifiable} because it predicts specific new 
particles \underline {necessarily} below the TeV energy scale.

\begin{table}[htb]
\caption{Values of $\alpha_s(M_Z)$ and $M_U/M_Z$ as functions of $M_S/M_Z$.}
\begin{center}
\begin{tabular}{|c|c|c|}
\hline 
$M_S/M_Z$ & $\alpha_s(M_Z)$ & $M_U/M_Z$ \\ 
\hline
1.0 & 0.119 & $2.1 \times 10^{14}$ \\ 
1.5 & 0.118 & $1.9 \times 10^{14}$ \\ 
2.2 & 0.117 & $1.7 \times 10^{14}$ \\ 
3.3 & 0.116 & $1.5 \times 10^{14}$ \\ 
5.0 & 0.115 & $1.4 \times 10^{14}$ \\ 
7.6 & 0.114 & $1.2 \times 10^{14}$ \\ 
\hline
\end{tabular}
\end{center}
\end{table}

The expected new particles and their supersymmetric partners are as follows. 
(1) Three copies of the exotic $h(h^c)$ quarks as in $SU(3)^3$ or $E_6$. 
(2) Eight axigluons \cite{fg87} corresponding to the breaking of $SU(3)_{CL} 
\times SU(3)_{CR}$ to $SU(3)_C$.  (3) Three sets of neutral $(3,3^*)$ $\eta$ 
particles which are reorganized into octets and singlets under $SU(3)_C$.  
(4) Three vector gauge bosons corresponding to the breaking of $SU(2)_{qL} 
\times SU(2)_{lL}$ to $SU(2)_L$.  (5) Three (2,2) bidoublets which are 
reorganized into triplets and singlets under $SU(2)_L$.  (6) Three sets 
of $SU(2)_L$ doublets $(\nu_x,e_x)$ and $(e_x^c,\nu_x^c)$.

The presence of $SU(2)_L$ nonsinglets is a potential phenomenological 
problem with respect to the precision electroweak measurements.  However, 
all such new particles, i.e. those listed under (4), (5), and (6) in the 
above, have masses which are invariant under $SU(2)_L$.  Hence their 
contributions to the electroweak oblique parameters are all very much 
suppressed.  They are however necessary because they render the $SU(2)_{qL} 
\times SU(2)_{lL}$ gauge extension \cite{lm03} anomaly-free.  This is also 
the purpose of the new $SU(3)_C$ octets and singlets (the $\eta$ particles) 
with respect to $SU(3)_{CL} \times SU(3)_{CR}$.  Thus this model has very 
unique predictions which are verifiable experimentally.

If the scale at which $SU(3)_{CL} \times SU(3)_{CR}$ reduces to the canonical 
$SU(3)_C$ is $M_A$ instead of $M_S$ with $M_A > M_S$, then 
Eqs.~(43) and (44) are modified to read
\begin{eqnarray}
{1 \over \alpha_s (M_Z)} &=& {3 \over 7} \left( {4 \over \alpha_2 (M_Z)} - 
{1 \over \alpha_Y (M_Z)} \right) + {1 \over 14 \pi} \left( 35 \ln {M_S 
\over M_Z} - 27 \ln {M_A \over M_Z} \right), \\ 
\ln {M_U \over M_Z} &=& {\pi \over 14} \left( {3 \over \alpha_Y (M_Z)} - 
{5 \over \alpha_2 (M_Z)} \right) - {1 \over 14} \left( 7 \ln {M_S \over M_Z} 
- 3 \ln {M_A \over M_Z} \right).
\end{eqnarray}
This would allow $M_S$ to be somewhat larger, say of order 1 TeV.

In conclusion, a new model of grand unification has been proposed based on 
$SU(3)^6$, which is a natural extension of two seemingly disparate notions, 
i.e. chiral color and quark-lepton nonuniversality.  In the context of 
supersymmetry, it has been shown that the five gauge couplings corresponding 
to $SU(3)_{CL} \times SU(3)_{CR} \times SU(2)_{qL} \times SU(2)_{lL} \times 
U(1)_Y$ naturally converge to a single value at $M_U$ of order $10^{16}$ GeV, 
and the scale at which this symmetry reduces to the standard $SU(3)_C \times 
SU(2)_L \times U(1)_Y$ is not more than a few times that of electroweak 
symmetry breaking.  This is a natural generalization of the MSSM and may be 
easily distinguished from it experimentally because specific new particles 
are predicted to exist below the TeV energy scale, and should be accessible 
at accelerators in the near future. \\[5pt]

This work was supported in part by the U.~S.~Department of Energy
under Grant No.~DE-FG03-94ER40837.

\newpage
\bibliographystyle{unsrt}

\end{document}